\documentclass[prb, reprint]{revtex4-1}
\pdfoutput=1
\usepackage{amssymb}
\usepackage{amsmath}
\usepackage{braket}
\usepackage{graphicx}
\usepackage{enumerate}
\usepackage[usenames,dvipsnames]{color}
\usepackage[colorlinks,bookmarks=false,citecolor=NavyBlue,linkcolor=OliveGreen,urlcolor=blue]{hyperref}
\newcommand{\be}{\begin{equation}}
\newcommand{\ee}{\end{equation}}
\newcommand{\ba}{\begin{aligned}}
\newcommand{\ea}{\end{aligned}}

\begin{document}
\title{Control of global properties in a closed many-body quantum system\\
by means of a local switch}
\author{Maurizio Fagotti}
\affiliation{D\'epartement de Physique, \'Ecole Normale Sup\'erieure/PSL Research University, CNRS, 24 rue Lhomond, 75005 Paris, France}
\begin{abstract}
We consider non-equilibrium time evolution after a quench of a global Hamiltonian parameter in systems described by Hamiltonians with local interactions. Within this background, we propose a protocol that allows to change global properties of the state by flipping a switch that modifies a local term of the Hamiltonian (creating a defect). A light-cone that separates two globally different regions originates from the switch. The expectation values of macroscopic observables, that is to say local observables that are spatially averaged within a subsystem, slowly approach new asymptotic values determined by the defect. The process is almost reversible: flipping again the switch produces a new light-cone with the two regions inverted. Finally, we test the protocol under repeated projective measurements. 
As explicit example we study the dynamics in a simple exactly solvable model but analogues descriptions apply also to generic models.  
\end{abstract}

\maketitle

In quantum mechanics the issue of locality/causality is very delicate and gave rise to brilliant debates extending over a century of quantum physics~\cite{EPR}. 
However in some contexts causality does emerge clearly. 
A shining example is given by the Lieb-Robinson bound~\cite{LR}, which states the existence of a maximal velocity at which information propagates in many-body systems with sufficiently fast decaying interactions.  The typical manifestation is the emergence of light-cones in the time evolution of local observables~\cite{LRpapers}. Importantly, over the time some hypotheses behind the original proof have been made milder~\cite{LRadv} and light-cones have been observed also in situations where, \emph{a priori}, a similar behavior would have not been expected~\cite{LRothers}. 

This kind of locality in quantum mechanics is solidly at the base of the descriptions of  subsystems at late time after global quenches in many-body quantum systems. 
The current understanding is that nonlocal conserved quantities do not affect the late time (stationary) expectation values (EVs) of local observables after a global Hamiltonian parameter is changed: the effective ensembles describing the late time physics can be obtained by maximizing the entropy under the constraints of the local and quasilocal conservation laws~\cite{GGEgeneral} (in quantum field theories the construction is more delicate~\cite{GGE-QFT}). This picture has the remarkable consequence that globally different ensembles describe the stationary properties of local observables evolving with Hamiltonians that look very similar to each other. 
A small integrability breaking perturbation is indeed  sufficient to reduce the infinite set of (quasi)local charges of the integrable model to a finite set, which in the most generic situation consists of just the Hamiltonian. The EVs of local observables experience in turn a crossover~\cite{Kollar,BEGR:preT, FB:super,FC:univentropy} from a quasi-stationary behavior, which can be captured by a generalized Gibbs ensemble~\cite{GGE0}, to the actual stationary value, describable instead by a Gibbs ensemble~\cite{GE}. These are usually known as prethermalization~\cite{QHubbard,metaOpt,Kollar,WSBK,BCK:KAM,E:preT,NIC:preT} or pre-relaxation~\cite{FB:super,FC:univentropy} behaviors. We stress that this picture applies also to the case where the unperturbed model is generic but there is at least a local charge besides the Hamiltonian. 

It is well known that also local perturbations can break global symmetries of a generic model or even integrability (in fact, a standard problem is to identify the local perturbations that preserve integrability~\cite{INTbc}). 

A (quasi)local conservation law of the unperturbed model (which can be either integrable or generic) can react to a local defect in two principal ways:
(i) it maintains its bulk part unchanged, 
(ii) it becomes extinct.

The first aim of this paper is to point out that the extinction of local charges results in a crossover between globally different states. Assuming local relaxation, this can be deduced from the time evolution of the EV of a charge $\tilde Q$ that becomes extinct. 
It turns out that the time derivative of the EV of $\tilde Q$ can \emph{not} be recast as a commutator between the Hamiltonian and some localized operator. Thus, generally the EV of the commutator remains nonzero also at late times and, in turn, the EV of $\tilde Q$ displays a typical linear behavior. One can then exhibit observables (essentially, the current associated with $\tilde Q$) with different EVs inside and outside a light-cone, even far away from the defect (see Sec. \ref{s:gen}): the light-cone separates regions with globally different properties.

We use this feature to engineer a local switch that allows one to change the state globally in an almost reversible way. 
The protocol is the following:
\begin{enumerate}
\item The system is prepared in the ground state $\ket{\Psi_0}$ of a translation invariant Hamiltonian $H_0$. 
\item At time $\tau_0$ a global parameter is changed and the state evolves with a new Hamiltonian $H_{\rm off}$ (global quench). Let us call $V$ a local perturbation that spoils a set of local conservation laws. A switch $\mathtt{s}$ is designed to turn $V$ on ($\mathtt{s}=\rm ON$) or off ($\mathtt{s}=\rm OFF$). 
\item After a sufficiently long time $T_1=\tau_1-\tau_0$ 
the switch is flipped (local quench $H_{\rm off}\rightarrow H_{\rm on}$).
\item At a later time $\tau_2$ the switch is flipped again (local quench $H_{\rm on}\rightarrow H_{\rm off}$), and so on. 
\end{enumerate}
We point out that similar dynamics were already considered \emph{e.g.} in Ref.~\cite{SM:oc}. 

\paragraph{Example.}

\begin{figure}[tbp]
\includegraphics[width=0.49\textwidth]{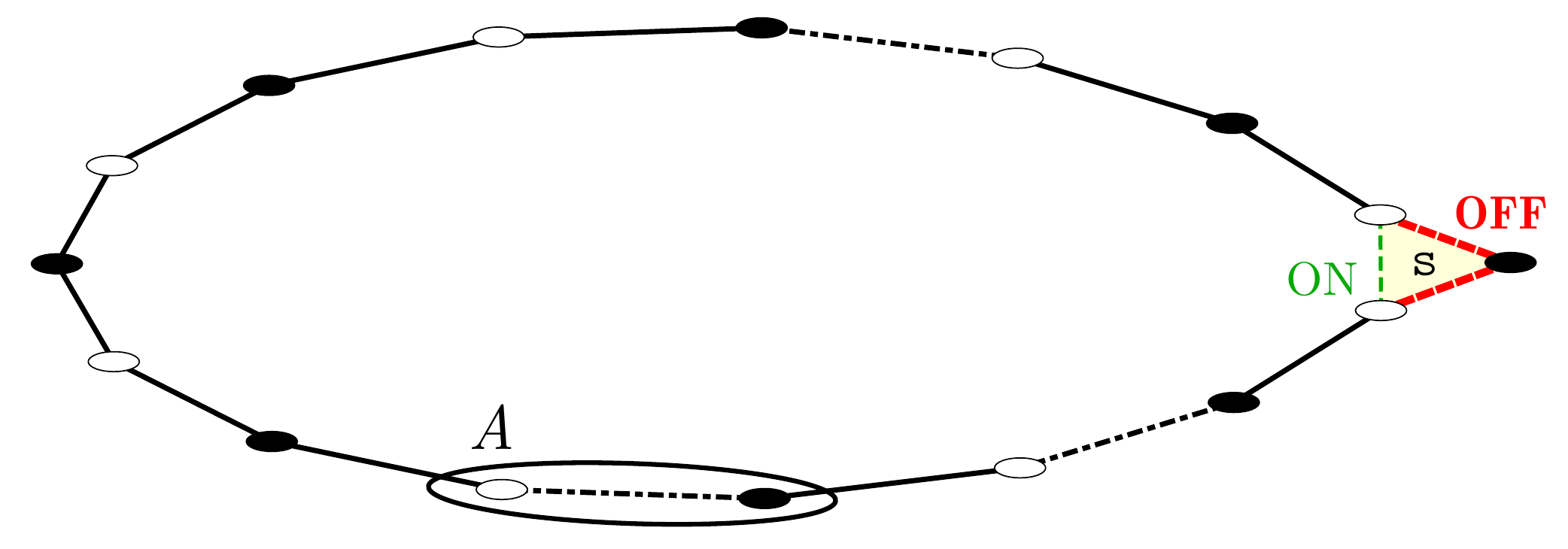}\\
\caption{Cartoon of the quench protocol. Full (empty) circles represent sites that are occupied (empty) at the initial time $\tau_0\equiv 0$ and the dot-dashed lines indicate an undefined number of sites. 
The switch $\mathtt{s}$ (light yellow triangle) decouples a site from the others.  
The (thick) red and (thin) green dashed lines represent connections when the switch is in the status of the corresponding color. The ellipse encircles the subsystem, $A$,  where macroscopic observables, like \eqref{eq:mzs}, are defined.
\label{f:cartoon}
}
\end{figure}
\begin{figure}[tbp]
\includegraphics[width=0.45\textwidth]{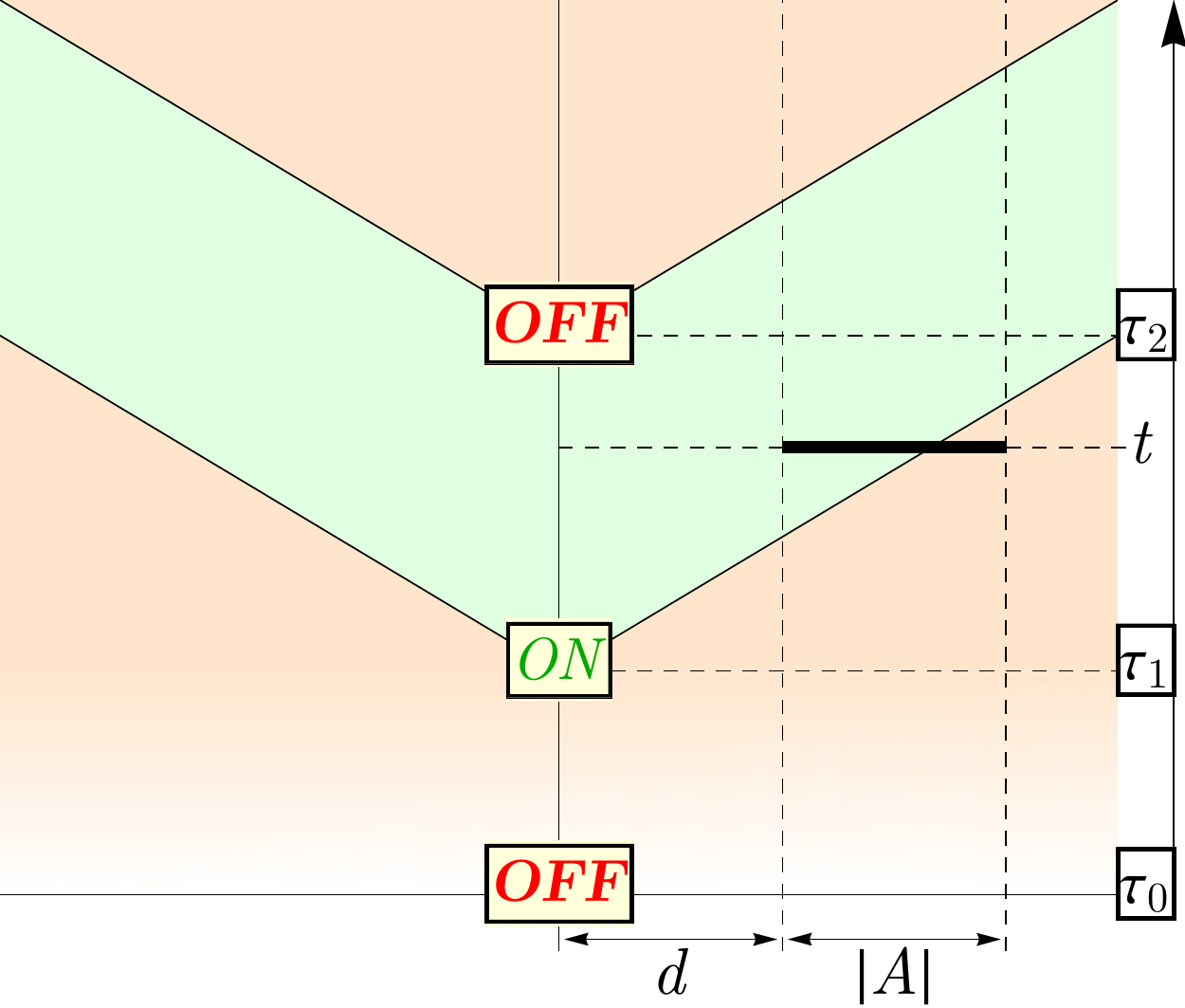}
\caption{Space-time picture of the light-cones propagating from the switch. Tags ``ON'' and ``OFF'' are placed at the times $\tau_i$ when the switch is turned on and off, respectively (before $\tau_1$ the state is two-site shift invariant). The thick segment  represents subsystem $A$ at time $t$. The subsystem is split in  parts belonging to different phases: in the green region EVs are approximately one-site shift invariant while in the orange region there is antiferromagnetic order. The crossover (orange to green) is still developing and the EVs of macroscopic observables vary almost linearly in time.  
\label{f:lightcone}
}
\end{figure}
For the sake of clarity, we analyze one of the simplest dynamics that can be controlled by a local switch through the mechanism described above and explained more extensively in Sec.~\ref{s:gen}. 
The reader can find further examples in Sec.~\ref{s:examples}. 
We consider a fermionic chain with Hamiltonian
\be\label{eq:H}
H_{\rm XY}=\sum_{\ell=1}^L\frac{1}{2}(c_\ell c^\dag_{\ell+1}-c^\dag_\ell c_{\ell+1})+\frac{\gamma}{2}(c_\ell c_{\ell+1}-c^\dag_\ell c^\dag_{\ell+1})\, ,
\ee
where $c_i$ satisfy the algebra $\{c_i^\dag,c_j\}=\delta_{i j}$, $\{c_i,c_j\}=0$ and periodic boundary conditions $c_{L+1}=c_1$ are imposed. In \eqref{eq:H} we omitted a multiplicative constant with the dimensions of an energy, which can be viewed as implicitly attached to the time. 
We note that Eq.~\eqref{eq:H} is mapped to the XY spin-$\frac{1}{2}$ chain~\cite{LSM:XY} by a Jordan-Wigner transformation. 
It was shown~\cite{FB:super} that this model has local conservation laws that break one-site shift invariance and are odd under a shift by one site. Clearly, they can exist  only if $L$ is even. Let us choose $L$ even and set the Hamiltonian $H_{\rm off}=H_{\rm XY}$. 
The chain's length can be effectively modified through a local perturbation. For example the following local term
\be\label{eq:V}
V=\frac{c^\dag_1(c_2+c_L+\gamma (c_2^\dag- c_L^\dag))-c^\dag_L c_2-\gamma c^\dag_L c^\dag_2}{2}+\mathrm{h.c.}
\ee
decouples the first site and has the practical effect of reducing the chain's length by one site. This in turn results in the extinction of infinitely many local conservation laws.  
We therefore set $H_{\rm on}=H_{XY}+V$. 
Since $V$ spoils only conservation laws that are odd under a shift by one site, we prepare the system in a state that breaks one-site shift invariance. In practice, this can be achieved by quenching from an antiferromagnetic or a dimerized phase.  In the rest of the paper the pre-quench state will be the N\'eel state (occupied sites alternating with empty sites). 
We point out that in the thermodynamic limit the distinction between odd and even chains is reduced to the presence or absence of a defect in the initial state at the position of the switch. 
By analogy with the phenomenon of metastability in phase transitions, the defect acts like a seed crystal from which a larger crystal will grow. 
Fig.~\ref{f:cartoon} sketches the system configuration. 

Globally different states can be distinguished by the EVs of macroscopic observables, \emph{i.e.} local observables that are averaged over a sufficiently large number of sites. We focus on the staggered magnetization per unit length in a connected subsystem $A$
\be\label{eq:mzs}
m^{z,s}[A]=-\frac{1}{|A|}\sum_{\ell\in A} (-1)^\ell \braket{c^\dag_j c_j}\, ,
\ee 
where $|A|$ is the subsystem's length, which we assume to be even. For even distances, at the initial time $m^{z,s}[A]=\frac{1}{2}$.
The time evolution of the N\'eel state under $H_{\rm off}$ can be easily worked out 
\be\label{eq:even}
\lim_{|A|\rightarrow\infty}\lim_{t\rightarrow\infty}m^{z,s}_{\rm off}[A]=\frac{1}{2}\frac{|\gamma|}{1+|\gamma|}\, .
\ee
On the other hand, time evolution under $H_{\rm on}$ results in a one-site shift invariant state (there are only one-site shift invariant local conservation laws), and hence
\be\label{eq:odd}
\lim_{|A|\rightarrow\infty}\lim_{t\rightarrow\infty}m^{z,s}_{\rm on}[A]=0 \, .
\ee
Our protocol of non-equilibrium dynamics is more complicated than a global quench but these limits are still appropriate reference values.

\paragraph{Analysis.}

The dynamics of macroscopic observables in $A$ are determined by four parameters:
\begin{enumerate}[(a)]
\item \label{e:chain}The chain's length $L$;
\item \label{e:sub}The subsystem's length $|A|$;
\item \label{e:dist}The distance $d$ between $A$ and the switch;
\item \label{e:delay}The time $T_i=\tau_{i}-\tau_{i-1}$ that elapses between consecutive changes of the switch status.
\end{enumerate}
The role of the parameters can be understood using a semiclassical picture~\cite{sc} based on the motion of quasiparticles excitations originated from/scattering with the switch. The quasiparticles travel at finite speed, bounded from above by the Lieb-Robinson velocity $v_M=||\gamma|-1|$. Therefore the light-cone emerging  from the switch propagates at $v_M$. Inside the light-cone a stationary state compatible with the perturbation is developing while, outside, the state time evolves as in the absence of the perturbation (\emph{cf.} Fig.~\ref{f:lightcone}).  Around the light-cone there is an unstable transient region.

\begin{figure}[tbp]
\includegraphics[width=0.45\textwidth]{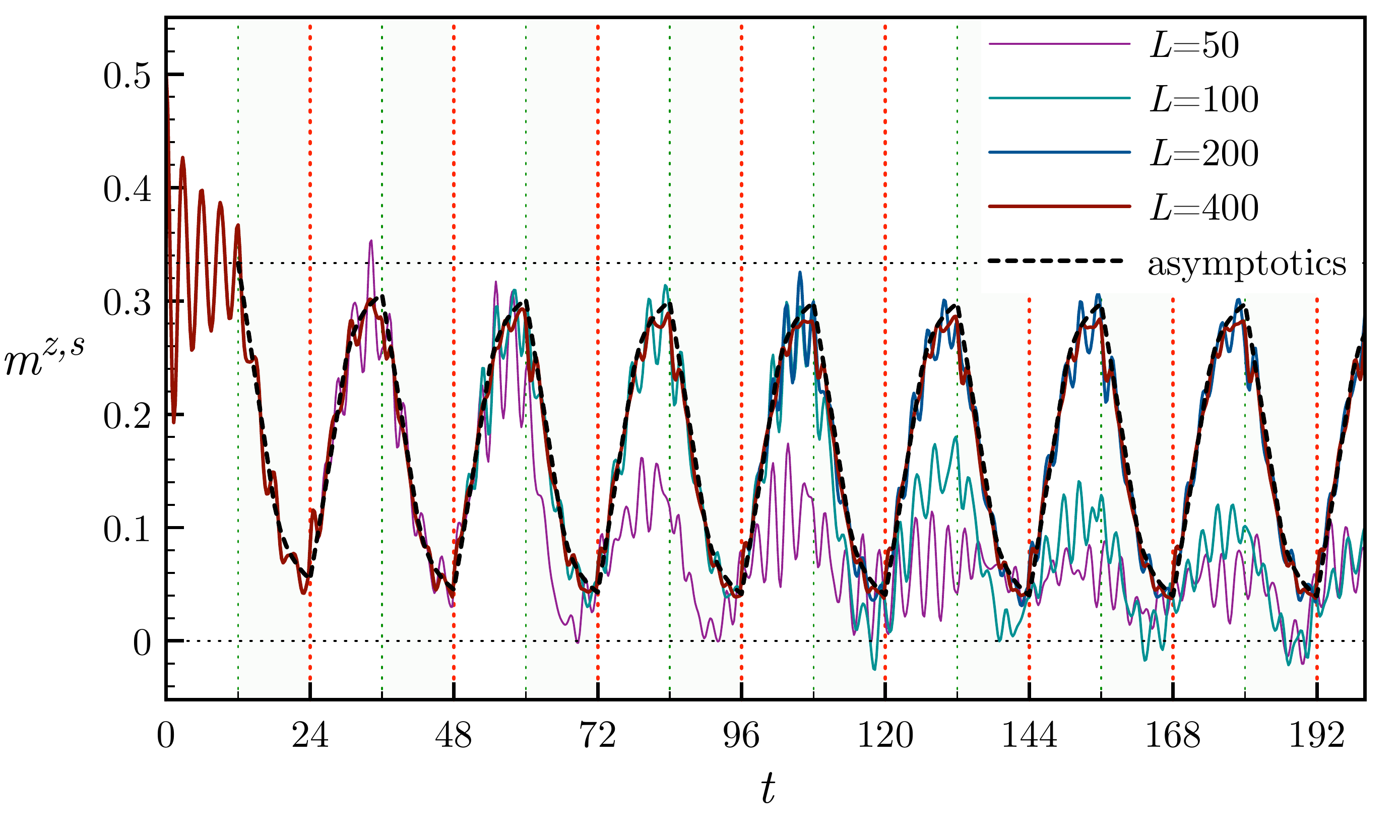}\\
\caption{The staggered magnetization per unit length in the subsystem consisting of the sites $\ell\in [2,7]$ ($|A|=6$, $d=0$) for $\gamma=2$ and various chain's lengths. The dotted vertical lines correspond to the times when the switch is activated (thin green) and deactivated (thick red); a faint green shadow highlights the intervals when the switch is on.
The dotted horizontal lines are the reference values \eqref{eq:even} and \eqref{eq:odd}. The dashed black curve is the prediction~\eqref{eq:prediction}. 
\label{f:mzsL}
}
\end{figure}

Our analysis shows that the finite size effects, \eqref{e:chain}, become visible after a time $t\sim L/(2 v_M)$ but the qualitative behavior is almost unchanged until a time $t\sim  L/v_M$ (see Fig.~\ref{f:mzsL}).  

\begin{figure}[tbp]
\includegraphics[width=0.45\textwidth]{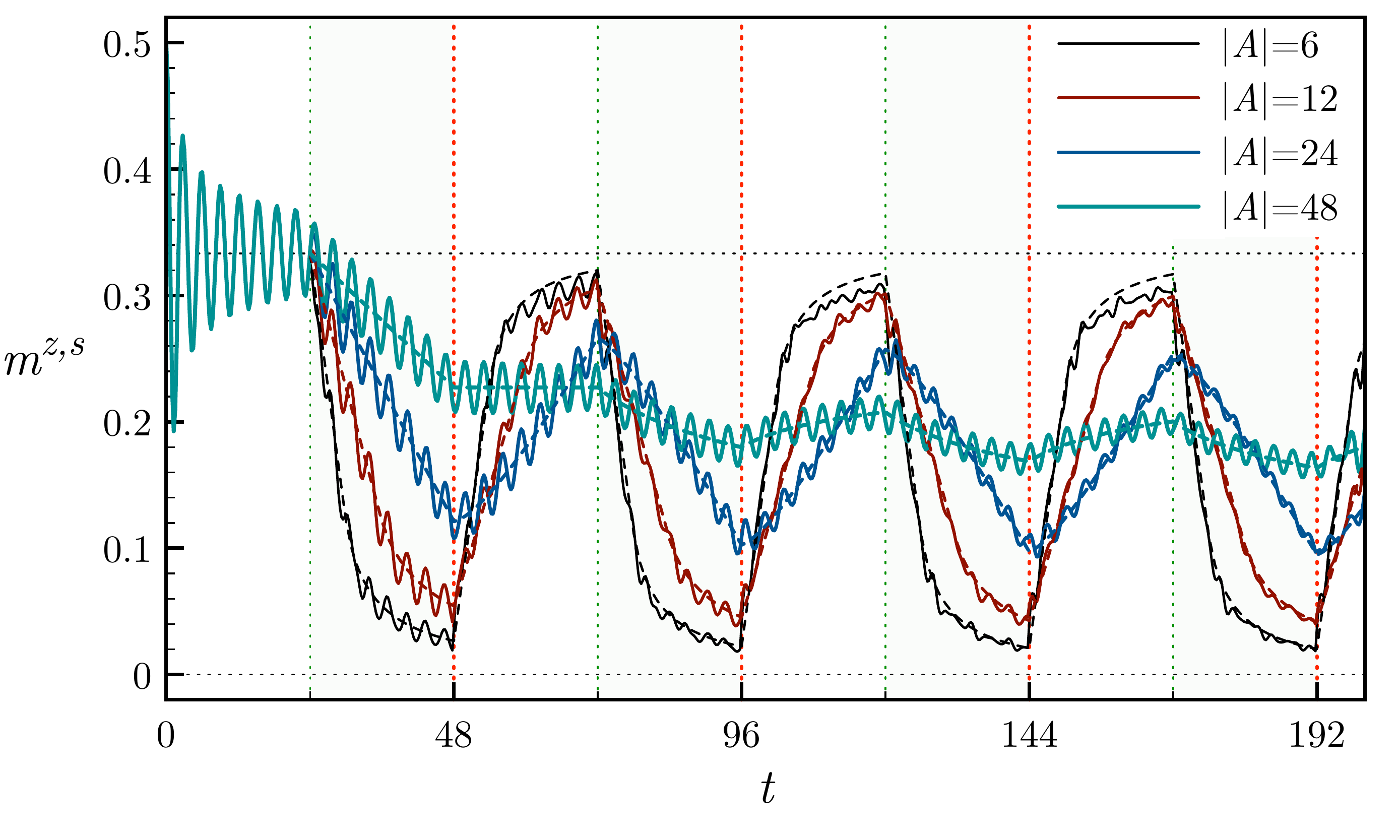}\\
\caption{The staggered magnetization per unit length for various subsystem's lengths starting from the second site ($d=0$) with the same quench parameters and notations of Fig.~\ref{f:mzsL} in a chain of $400$ sites. The dashed lines are the predictions~\eqref{eq:prediction}. 
\label{f:mzsA}
}
\end{figure}

The main effect related to the subsystem's length, \eqref{e:sub}, is in the time that the light-cone employes to flow through the subsystem: the larger $A$ is and the slower the transition between one state and the other (\emph{cf}.~Fig.~\ref{f:mzsA}). During this time the EVs of macroscopic observables experience a typical linear time evolution: the light-cone, moving at the Lieb-Robinson velocity, splits the subsystem in parts that approximately belong to different phases.

\begin{figure}[tbp]
\includegraphics[width=0.45\textwidth]{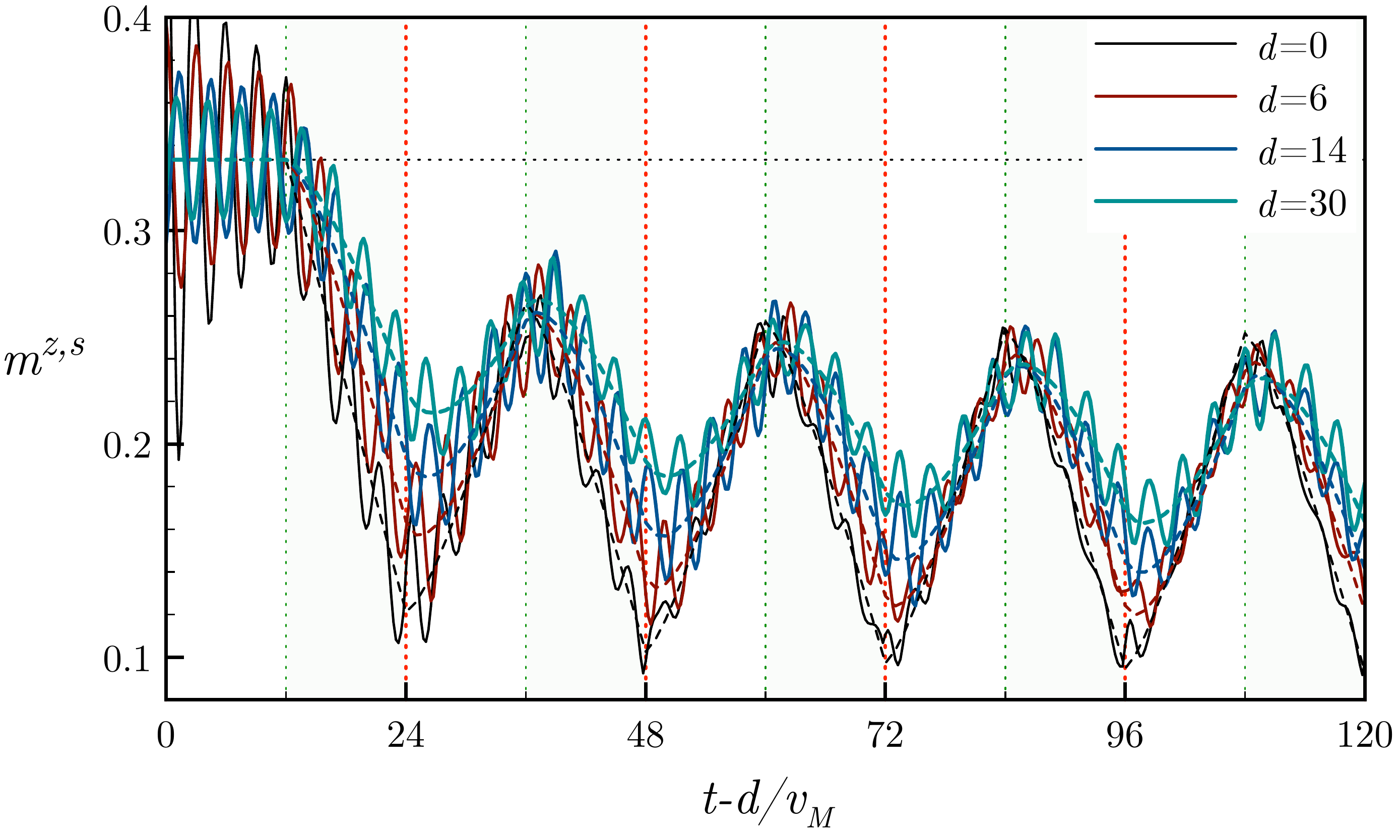}\\
\caption{The staggered magnetization per unit length in the subsystem $[2+d,13+d]$ with the same quench parameters and notations of Fig.~\ref{f:mzsL} in a chain of $400$ sites for various distances $d$. The dashed lines are the predictions~\eqref{eq:prediction}. 
The time is shifted to compensate for the finite velocity at which information propagates.
\label{f:mzsd}
}
\end{figure}

 The distance $d$ from the switch, \eqref{e:dist}, introduces a delay $\delta t=d/v_{M}$ between the time at which the switch is flipped and the time at which  observables in $A$ start feeling the change. In addition,  only the quasiparticles with velocity $v\geq v_M /(1+\frac{v_M (t-\delta t)}{d})$ have reached the subsystem, so the transition is also slower at larger distances, where the effects of the unstable regions around the light-cones are amplified (see Fig.~\ref{f:mzsd}).  
 
 \begin{figure}[tbp]
\includegraphics[width=0.45\textwidth]{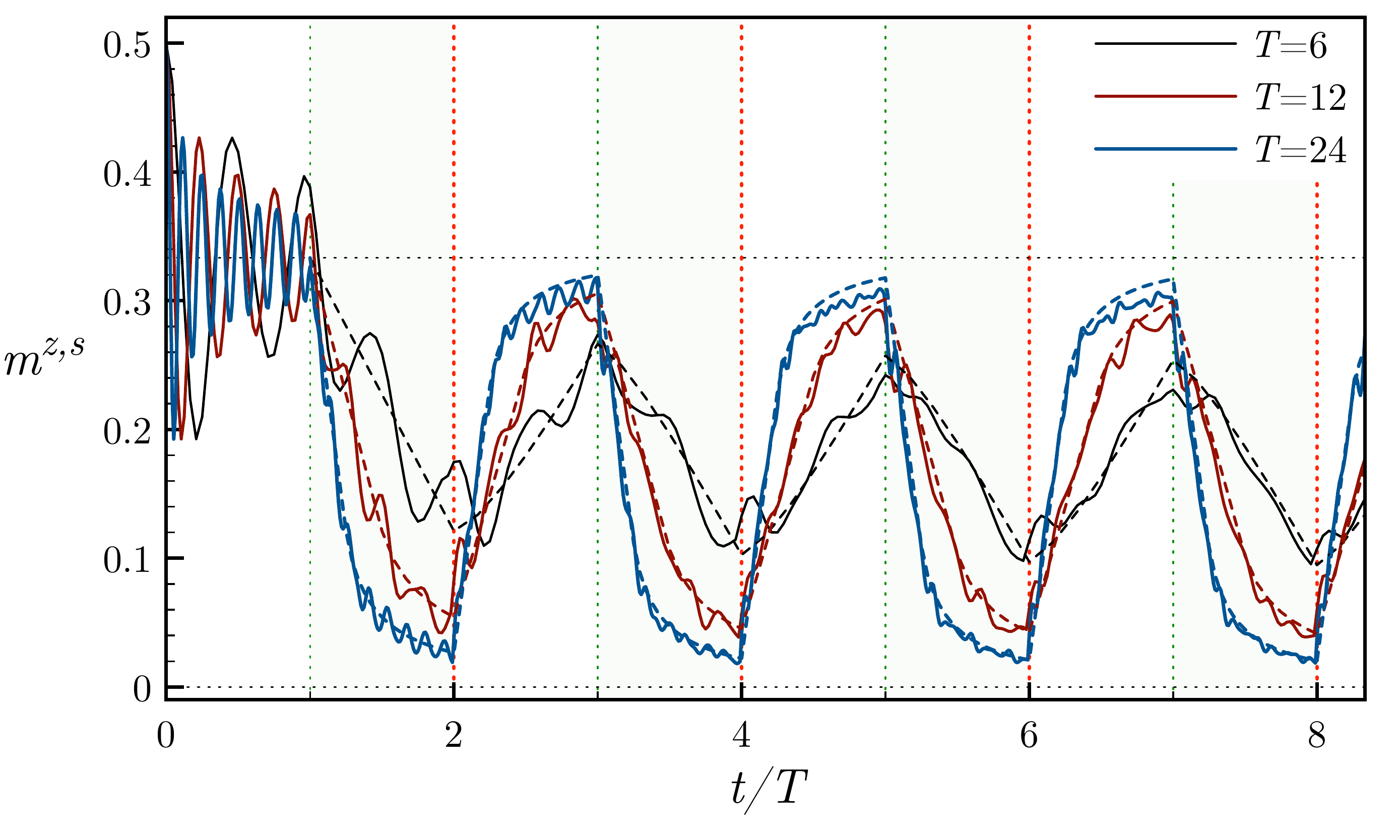}\\
\caption{The staggered magnetization per unit length in the subsystem $[2,7]$ ($|A|=6$, $d=0$) with the same quench parameters and notations of Fig.~\ref{f:mzsL} in a chain of $400$ sites for various time intervals $T_i=T$ between successive switches. 
The dashed lines are the predictions~\eqref{eq:prediction}. 
\label{f:mzsT}
}
\end{figure}
 
The time $T_i$  between consecutive local quenches, \eqref{e:delay}, puts a limit on 
 the time left for the subsystem to respond to the change (\emph{cf}. Fig.~\ref{f:mzsT}).
 
Following a semi-quantitative reasoning based on the semiclassical picture of Fig.~\ref{f:lightcone}, we conjecture an asymptotic expression for $m^{z,s}$ (see Sec.~\ref{s:semi}), which is expected to hold in the thermodynamic limit when all the parameters are sufficiently large:
\be\label{eq:prediction}
m^{z,s}\sim \frac{1}{2}\frac{|\gamma|}{1+|\gamma|}+\int_{\pi}^\pi\frac{\mathrm d k}{2\pi} \frac{\gamma^2\sin^2 (k/2)\mu^{A}_{t,\{\tau\}}(k)}{1+\cos k+\gamma^2(1-\cos k)}\, .
\ee
Here $\{\tau\}$ is the set of the times $\tau_j$ at which the switch is flipped; the details of the dynamics are encoded in a single function
\begin{multline}\label{eq:mu}
\mu^{A}_{t,\{\tau\}}(k)=\frac{1}{|A|}\sum_{j=1}(-1)^{j}\theta(t-\tau_j)\times\\
\bigl[\min\bigl(2|\varepsilon'_k|(t-\tau_j),|A|+d\bigr)-\min\bigl(2|\varepsilon'_k|(t-\tau_j),d\bigr)\bigr]
\end{multline}
where $\varepsilon_k=(\cos^2\frac{k}{2}+\gamma^2\sin^2\frac{k}{2})^{1/2}$ is the dispersion relation of the two species of quasiparticles that diagonalize $H_{\rm off}$ \eqref{eq:H} with the momentum $k$ defined in terms of translations by multiples of two sites.
The numerical data are in good agreement with \eqref{eq:prediction}, which indeed captures the relevant features of the dynamics in all the situations investigated (\emph{cf}. Figs \ref{f:mzsL}, \ref{f:mzsA}, \ref{f:mzsd}, \ref{f:mzsT}).  

Although the quench details affect the time evolution of macroscopic observables (as a consequence of the transitions being always incomplete), the entire process shows surprising reversible features: the profile of the staggered magnetization is only weakly influenced by the past actions on the switch. 
This can be inferred also from \eqref{eq:prediction} and \eqref{eq:mu}: when $t\gg \tau_{\bar j}$ the contribution to $\mu_{t,\{\tau\}}^A(k)$ from the switches at $\tau_j\leq \tau_{\bar j}$ is reduced to $\sum_{j\leq \bar j}(-1)^{j}$, which has the practical effect of resetting the EV to \eqref{eq:even}, for $\bar j$ even, or \eqref{eq:odd}, for $\bar j$ odd.

\paragraph{Projective measurement.}
\begin{figure}[!t]
\includegraphics[width=0.49\textwidth]{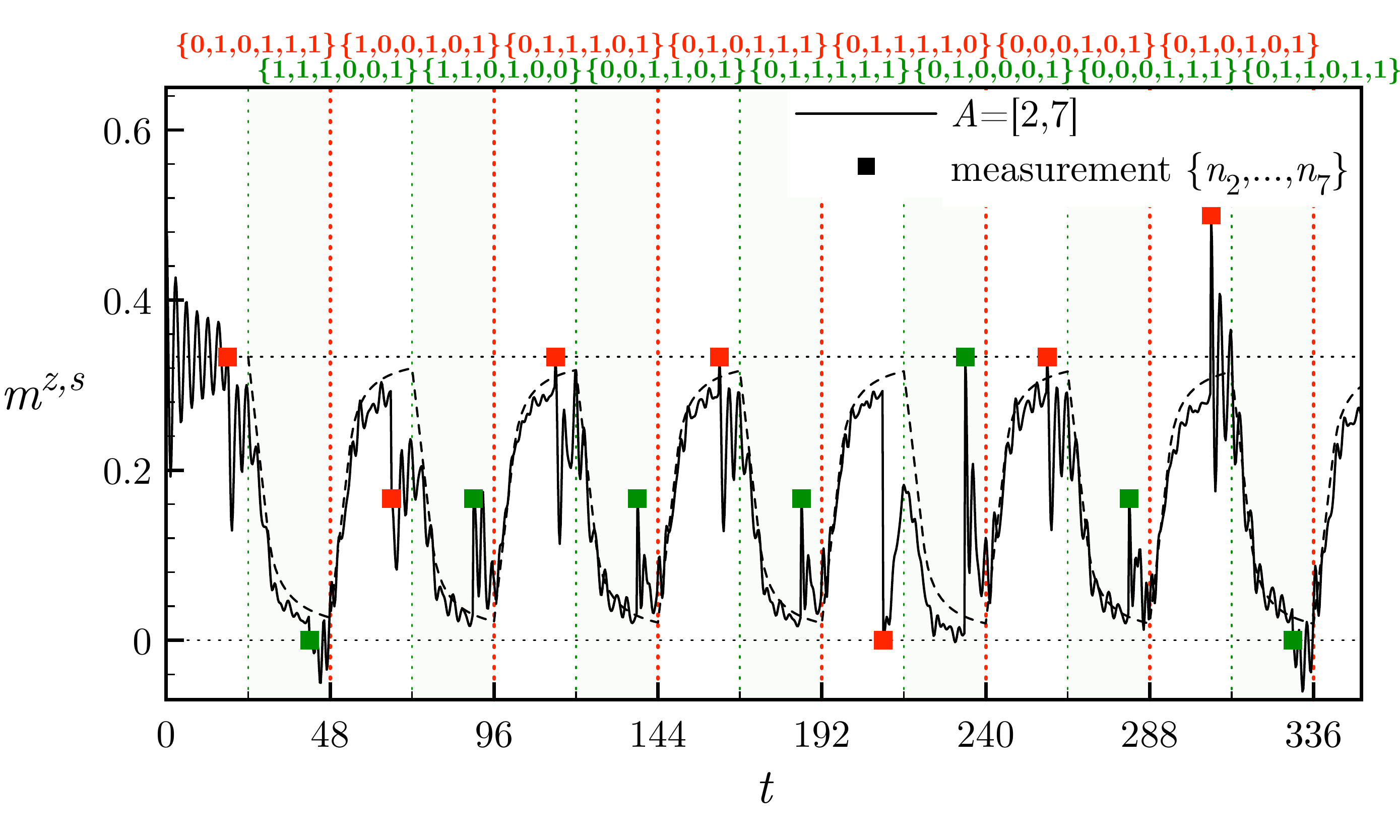}
\caption{The staggered magnetization per unit length in the subsystem $[2,7]$ ($|A|=6$, $d=0$) with the same quench parameters and notations of Fig.~\ref{f:mzsL} in a chain of $800$ sites. 
The squares (red, when $\mathtt{s}=$ OFF, and green, otherwise) represent (simultaneous) projective measurements of all the occupation numbers in $A$, repeated at time intervals of $24$ starting from $t=18$ (the outcomes are shown on top with the  color of the squares). The dashed curve is \eqref{eq:prediction} without the measurements.  
\label{f:mzsM}
}
\end{figure}
Here we investigate whether our construction is stable under projective measurements (see Sec.~\ref{s:proj}). 
Fig.~\ref{f:mzsM} shows $m^{z,s}$ for $|A|=6$ following repeated measurements of  all the occupations numbers $n_j=c_j^\dag c_j$ in $A$. Generally, after the measurement there is a time interval characterized by sizable fluctuations that make the time evolution noisier. Nevertheless the qualitative behavior of $m^{z,s}$ does not change. In average the measurements distinguish the two phases.

\paragraph{Summary and discussion.}
We proposed a general protocol of non-equilibrium  evolution that could be used to engineer a quantum device in which global properties of a state are controlled by a local switch. 
We tested the protocol on a quantum XY fermionic chain, where we
analyzed the staggered magnetization averaged in a subsystem and conjectured an asymptotic expression  
for its time evolution. In order to assess the stability of the setup we kept all the parameters finite and tested the protocol under projective measurements.

The effect described relies on the absence of a one-to-one correspondence between  (the bulk part of) the local conservation laws with and without a local perturbation. 
These situations are very common and emerge also in generic models with a global symmetry (\emph{e.g.} particle number conservation) if the switch is associated with a symmetry-breaking defect (see Sec.~\ref{s:examples} for an explicit example).  The generality of this situation opens the door to experimental realizations. 

In integrable models the switch could change the boundary conditions  from periodic to open, where the set of local charges is almost halved \cite{GM:open, F:XYopen} (see also Sec.~\ref{s:examples}). 
More general local perturbations would spoil integrability, as could be inferred also from Ref.~\cite{T-HS}. 

Finally, we note some similarities with the protocol of joining together two chains to form a single one (and other similar dynamics), considered \emph{e.g.} in Ref.~\cite{joinchainquench}.
Also in that case a light-cone that separates globally different states is originated from a given point (the junction). 
The two situations contrast in the initial states, which we assumed to be homogeneous and non-stationary in the bulk.

\begin{acknowledgments}
I thank Fabian Essler for helpful discussions and Bruno Bertini for useful comments.
This work was supported by the LabEx ENS-ICFP: ANR-10-LABX-0010/ANR- 10-IDEX-0001- 02 PSL*.
\end{acknowledgments}

\onecolumngrid
%\newpage
\section*{Supplemental Material}
\section{Localized defects in spin chain Hamiltonians}\label{s:gen}
Let us indicate with $H_0$ a generic homogeneous Hamiltonian with periodic boundary conditions and with $H_1$ the same Hamiltonian with a defect $d(r_0)$ localized around position $r_0$:
\be
H_1=H_0+d(r_0)\, .
\ee 
We consider homogenous initial states $\ket{\Psi_0}$ and focus on situations where the expectation values of local observables relax to stationary values both under $H_0$ and under $H_1$.

An important consequence of local relaxation is that the expectation value of the commutator between the Hamiltonian and any operator $\mathcal O$ with a finite typical range approaches zero in the limit of infinite time, indeed
\be
\partial_t\braket{\Psi_0|e^{i H_{\mathtt s} t}\mathcal Oe^{-i H_{\mathtt s} t}|\Psi_0}\rightarrow 0\Rightarrow \braket{\Psi|e^{i H_{\mathtt s} t}i[H_{\mathtt s},\mathcal O]e^{-i H_{\mathtt s} t}|\Psi_0}\rightarrow 0\, .
\ee
Let us now consider a conservation law $Q$ of $H_0$ that can be deformed close to the defect so as to be conserved also for $\mathtt s=1$:
\be\label{eq:norm}
[Q,H_0]=0\qquad [Q+\delta Q,H_1]=0\qquad \parallel\delta Q\parallel<\infty\, ,
\ee
where $\parallel\cdot\parallel$ is the operator norm (the maximal eigenvalue in absolute value).  

The variation of the time evolution of $Q$ under $H_1$ is given by
\be\label{eq:bound}
|\braket{\Psi_0|e^{i H_1 t}Q e^{-i H_1 t}-Q|\Psi_0}|=|\braket{\Psi_0|\delta Q-e^{i H_1 t}\delta Qe^{-i H_1 t}|\Psi_0}|\leq 2\parallel \delta Q\parallel \, .
\ee
Being the norm of $\delta Q$ finite (see \eqref{eq:norm}), the expectation value of $Q$, which is generally proportional to $L$, remains close to the original value at any time after the quench.

In the physical situations when the infinite time limit of the expectation values of local observables commutes with the limit of large distance from the defect, it is reasonable to expect that the stationary behavior of observables can be described by the statistical ensemble that maximizes the entropy under the constraints of the integrals of motion (this is indeed the principle that holds in the limit $\lim_{t\rightarrow\infty}\lim_{d\rightarrow\infty}$). In all the cases studied we found that this happens when all the conservation laws that are relevant to the stationary properties for $\mathtt s=0$ are simply deformed by the defect
\be
Q_j^{(\mathtt s=1)}= Q_j^{(\mathtt s=0)}+\delta Q_j\, .
\ee 

On the other hand, the situation becomes much more involved if there are charges of $H_0$ that become extinct. For example, let us assume that there is a charge $\tilde Q$ of $H_0$ such that 
\be\label{eq:A}
\mathcal A_{\tilde Q}\overset{def}{=}\lim_{t\rightarrow\infty}\braket{\Psi_0|e^{i H_1 t}i [d(r_0),\tilde Q]e^{-i H_1 t}|\Psi_0}\neq 0
\ee
Since
\be
[H_1,\tilde Q]=[H_0+d(r_0),\tilde Q]=[d(r_0),\tilde Q]\, ,
\ee
the variation of the expectation value of $\tilde Q$ is given by
\be
|\braket{\Psi_0|e^{i H_1 t}\tilde Q e^{-i H_1 t}-\tilde Q|\Psi_0}|=
|\int_0^t\mathrm d \tau \braket{\Psi_0|e^{i H_1 \tau}i[d(r_0),\tilde Q] e^{-i H_1 \tau}|\Psi_0}|\, .
\ee
The right hand side can be bounded from below as follows:
\begin{multline}
|\int_0^t\mathrm d \tau \braket{\Psi_0|e^{i H_1 \tau}i[d(r_0),\tilde Q] e^{-i H_1 \tau}|\Psi_0}|\\
=|\int_0^T\mathrm d \tau \braket{\Psi_0|e^{i H_1 \tau}i[d(r_0),\tilde Q] e^{-i H_1 \tau}|\Psi_0}+\int_T^t\mathrm d \tau \braket{\Psi_0|e^{i H_1 \tau}i[d(r_0),\tilde Q] e^{-i H_1 \tau}|\Psi_0}|\\
\geq -T\parallel i[d(r_0),\tilde Q] \parallel +(t-T)\min_{\tau>T}| \braket{\Psi|e^{i H_1 \tau}i[d(r_0),\tilde Q] e^{-i H_1 \tau}|\Psi}|\, ,
\end{multline}
where $T$ is sufficiently large that
\be
\min_{\tau>T}| \braket{\Psi_0|e^{i H_1 \tau}i[d(r_0),\tilde Q] e^{-i H_1 \tau}|\Psi_0}|>0\, .
\ee
Since $\parallel i[d(r_0),\tilde Q] \parallel$ is finite (the operator is quasilocalized around the defect), this gives a lower bound to  the variation of the expectation value of $\tilde Q$ that increases with the time, diverging in the limit of infinite time. By \eqref{eq:bound}, there can not be a corrective term $\delta \tilde Q$ that satisfies \eqref{eq:norm}, and hence there is no quasilocal deformation of $\tilde Q$ that makes the operator commuting with $H_1$: $\tilde Q$ is destroyed by the defect. 
A linear increase (or decrease) of the expectation value of a charge of $H_0$ turns out to be the typical signature of its extinction.

We now show that a nonzero $\mathcal A_{\tilde Q}$, \eqref{eq:A}, results in the propagation of a light-cone that separates regions with globally different properties. 
To that aim we consider the density $\tilde q_\ell$ of the extinct charge $\tilde Q$, where index $\ell$ signifies that the operator acts nontrivially only around site $\ell$
\be
\tilde Q=\sum_\ell \tilde q_\ell\, .
\ee 
Since $\partial_t\braket{\Psi_t|\tilde q_\ell|\Psi_t}=\braket{\Psi_t|i[H,\tilde q_\ell]|\Psi_t}$,
summing over the sites in an interval $[d+r_0,n+r_0]$, with $0<d<n$, gives the continuity equation
\be
\partial_t\braket{\Psi_t|\sum_{\ell=d+r_0}^{n+r_0}\tilde q_\ell|\Psi_t}=\braket{\Psi_t|i[H_1,\sum_{\ell=d+r_0}^{n+r_0}\tilde q_\ell]|\Psi_t}=\braket{\Psi_t|J_{n+r_0}-J_{d+r_0}|\Psi_t}\, .
\ee
Here we used that the extinct charge is conserved in the bulk, and hence the commutator can only depend on surface terms, which we called $J$ (they are local operators). We note that, as long as $d$ is larger than the sum of the ranges of $H_0$ and $\tilde Q$, the definition of $J$ does not change if $H_1$ is replaced by $H_0$. 
Let us now consider the limit $d\ll v t\ll n$, with $v$ the Lieb-Robinson velocity. We find
\begin{multline}
\frac{\partial}{\partial t}\braket{\Psi_t|\sum_{d\leq |\ell-r_0|\leq n}\tilde q_\ell|\Psi_t}=\frac{\partial}{\partial t}\braket{\Psi_t|\tilde Q|\Psi_t}-\frac{\partial}{\partial t}\braket{\Psi_t|\sum_{|\ell-r_0|<d}\tilde q_\ell|\Psi_t}-\frac{\partial}{\partial t}\braket{\Psi_t|\sum_{|\ell-r_0|>n}\tilde q_\ell|\Psi_t}\\
\xrightarrow{v t\ll n} \braket{\Psi_t|i[d(r_0),\tilde Q]|\Psi_t}-\frac{\partial}{\partial t}\braket{\Psi_t|\sum_{|\ell-r_0|<d}\tilde q_\ell|\Psi_t}\xrightarrow{d\ll vt} \mathcal A_{\tilde Q}\, .
\end{multline}
In the first step of the second line we used that the density is practically constant until the information about the defect has not yet reached the site; in the second step we instead used that, at times $t\gg d/v$, local observables relax. 
Putting all together we get
\be\label{eq:EVcurr}
\braket{\Psi_t|(J_{n+r_0}-J_{d+r_0})+(J_{r_0-d}-J_{r_0-n})|\Psi_t}\xrightarrow{d\ll v t\ll n}\mathcal A_{\tilde Q}\neq 0\, ,
\ee
where, we stress, $J_{n+r_0}$ and $J_{r_0-n}$ ($J_{d+r_0}$ and $J_{r_0-d}$) are outside (inside) the light-cone. 
This means that the expectation value of $J_\ell$ depends on whether $\ell$ is inside or outside the light-cone, even for $d\gg 1$.

One can use similar arguments to show the effects of flipping the switch.
For example, let us assume to turn off the switch at time $\tau_2$. Because of the Lieb-Robinson bounds, at times $t\lesssim \tau_2+d/v$ the expectation value of $J_{d+r_0}$ (and, in turn, of $J_{n+r_0}$) can not be changed. 
On the other hand, in the opposite limit $d\ll n\ll v(t-\tau_2)+d \ll vt$, it is reasonable to expect relaxation to the values corresponding to time evolution under $H_0$, which in fact are the initial values.  
Let us then consider the remaining two nontrial limits  
\begin{enumerate}
\item \label{e:a} $d\ll v(t-\tau_2)+d \ll vt\ll n$;
\item \label{e:b} $d\ll v(t-\tau_2)+d \ll n\ll vt$.
\end{enumerate}
In case \ref{e:a} we find
\be
\frac{\partial}{\partial t}\braket{\Psi_t|\sum_{d\leq |\ell-r_0|\leq n}\tilde q_\ell|\Psi_t}
\xrightarrow{v t\ll n} \braket{\Psi_t|i[d(r_0),\tilde Q]|\Psi_t}-\frac{\partial}{\partial t}\braket{\Psi_t|\sum_{|\ell-r_0|<d}\tilde q_\ell|\Psi_t}\xrightarrow{d\ll v\tau_2} 0\, ,
\ee
where we used that now $\tilde Q$ is conserved and hence $\braket{\Psi_t|i[d(r_0),\tilde Q]|\Psi_t}\rightarrow 0$. This results in 
\be\label{eq:EVcurra}
\braket{\Psi_t|(J_{n+r_0}-J_{d+r_0})+(J_{r_0-d}-J_{r_0-n})|\Psi_t}\xrightarrow{d\ll v(t-\tau_2)+d \ll vt\ll n}0\, .
\ee
Since $\braket{J_{n+r_0}}$ had still the initial value, this suggests that also $J_{d+r_0}$ returned to the original value. 

Case \ref{e:b} can be deduced from the other limits: $J_{n+r_0}$ turns out to have the value corresponding to $H_1$, whereas $J_{d+r_0}$ already relaxed to the initial value. 

We stress that we did not use any specific property of the Hamiltonian, so this physical picture applies to noninteracting, interacting integrable and even generic models with at least a local charge besides the Hamiltonian. 

In conclusion, in the presence of extinct charges we should expect that the stationary state that describes the late time properties under $H_1$ deep inside the light-cone is macroscopically different from the stationary state that gives the correct behavior outside the light-cone (which can be simply obtained replacing $H_1$ by $H_0$). In other words, the limit of infinite time does not commute with the limit of infinite distance.
In addition, flipping the switch corresponds to reverse the situation. 

\section{Semiclassical formula for (simple) macroscopic observables}\label{s:semi}
Let us focus on the model \eqref{eq:H}. 
The space-time picture of Fig.~\ref{f:lightcone} can be easily translated into a semiclassical formula for the time evolution of macroscopic observables. 
We consider extensive quadratic operators $\mathcal O^{-}$ that are \emph{odd} under a shift by one site and restrict them to a subsystem $A$ as follows
\be\label{eq:Oodd}
\mathcal O^{-}_A=\frac{1}{|A|}\frac{\mathrm{tr}_{\bar A}[\mathcal O^{-}]\otimes \mathrm I_{\bar A}}{\mathrm{tr}_{\bar A}[\mathrm I_{\bar A}]}\, ,
\ee 
where $\bar A$ is the complement of $A$. The staggered magnetization per unit length $m^{z,s}$ considered in the main text belongs to this class.

If the time $\tau_1$ of the first switch is sufficiently large, we can assume $\braket{\mathcal O^{-}_A}$ to have approximately reached the GGE value associated with time evolution under $H_{\rm off}$; this is an integral over the momentum (defined in terms of translations by multiples of two sites) of a function depending on the observable
\be
\braket{\mathcal O^{-}_A}^{\rm off}_\infty=\lim_{t\rightarrow\infty}\braket{\Psi_0|e^{iH_{\rm off} t}\mathcal O^{-}_Ae^{-iH_{\rm off} t}|\Psi_0}= \int_{-\pi}^\pi\frac{\mathrm d k}{2\pi} f(k; \mathcal O^-)\, .
\ee 
We interpret $f(k; \mathcal O^-)$ as the contribution given by the quasiparticle with momentum $k$ to the expectation value. 
In particular, for the staggered magnetization we find
\be
f(k; m^{z,s})=\frac{\gamma^2 \sin^2(k/2)}{1+\cos k+\gamma^2(1-\cos k)}\, .
\ee
On the other hand, under $H_{\rm on}$ the expectation values of odd observables approach zero. 

The first time that the switch is flipped ($\mathtt s\rightarrow$ ON) a destructive light-cone emerges. This is responsible for a negative contribution $-f(k;\mathcal O^-)$ when the quasiparticle is (semiclassically) inside $A$. Flipping again the switch the new light-cone is instead associated with positive contributions $f(k;\mathcal O^-)$, and so on. This results in
\be\label{eq:scO}
\braket{\Psi_t|\mathcal O^{-}_A|\Psi_t}\sim  \int_{-\pi}^\pi\frac{\mathrm d k}{2\pi} f(k; \mathcal O^-)\Bigl[1+\mu^{A}_{t,\{\tau\}}(k)\Bigr]\, ,
\ee
where $\mu^{A}_{t,\{\tau\}}(k)$, given by
\be\label{eq:mu}
\mu^{A}_{t,\{\tau\}}(k)=\frac{1}{|A|}\sum_{j=1}(-1)^{j}\theta(t-\tau_j)\bigl[\min\bigl(2|\varepsilon'_k|(t-\tau_j),|A|+d\bigr)-\min\bigl(2|\varepsilon'_k|(t-\tau_j),d\bigr)\bigr]\, ,
\ee
counts the number of quasiparticles with momentum $k$ in $A$ (per unit length) with an overall sign depending on the status of the switch when the quasiparticles were produced. 

For quadratic operators that are \emph{even} under a shift by one site, the semiclassical approach does not show any effect caused by the switch. 
We do not expect that this holds true for non-quadratic (even) operators, since the decomposition of their expectation values (EVs) by Wick's theorem involves also EVs of operators that are odd under a shift by one site.

\section{Projective measurements}\label{s:proj}
A (von Neumann) projective measurement $\mu[\mathcal O]$ of an observable $\mathcal O$ involves the collapse of the wave function
\be
\ba
&\mathcal O=\sum_i \lambda_i P_{\lambda_i}\qquad  (P_{\lambda_i}P_{\lambda_j}=\delta_{i j}P_{\lambda_i},\ \lambda_i=\lambda_j\Leftrightarrow i=j)\\
& \ket{\Psi}\xrightarrow{\mu[\mathcal O]=\lambda_{\bar i}} \frac{P_{\lambda_{\bar i}}\ket{\Psi}}{\sqrt{\braket{\Psi|P_{\lambda_{\bar i}}|\Psi}}}\qquad p(\mu[\mathcal O]=\lambda_i)=\braket{\Psi|P_{\lambda_{\bar i}}|\Psi}\, ,
\ea
\ee  
where $p(\mu[\mathcal O]=\lambda)$ is the probability of measuring $\lambda$. 
In principle, the collapse could disrupt the nice local properties that characterize our protocol of non-equilibrium time evolution. The response of the system to projective measurements is therefore a severe test for the dynamics considered in this paper. 

Since the Hamiltonian is quadratic and the initial state is a Slater determinant, we can restrict ourselves to pre-measurement states $\ket{\Psi}$ that are Slater determinants.
However, after a measurement the state becomes generally very complicated anyway.   
For the sake of simplicity we consider measurements that preserve gaussianity, 
a sufficient condition being the applicability of  Wick's theorem to the density matrix $P_{\lambda_j}/\mathrm{tr}[P_{\lambda_j}]$, for any $j$. 
A simple observable with  this property is the occupation number $n_j=c^\dag _j c_j$. Its measurement results in the following collapse of the density matrix
\be
\ket{\Psi}\bra{\Psi}\rightarrow \left\{\begin{aligned}
&\frac{n_j \ket{\Psi}\bra{\Psi}n_j}{\braket{\Psi|n_j|\Psi}}&& (p=\braket{\Psi|n_j|\Psi})&& \mu[n_j]=1\\
&\frac{(1-n_j)\ket{\Psi}\bra{\Psi}(1-n_j)}{1-\braket{\Psi|n_j|\Psi}}&& (p=1-\braket{\Psi|n_j|\Psi})&& \mu[n_j]=0\, .
\end{aligned}
\right.
\ee
Any expectation value can be written in terms of the two-point correlation functions, which are conveniently written in terms of the Majorana fermions ($\{a_i, a_j\}=2\delta_{i j}$)
\be\label{eq:a}
a_{2i-1}=c^\dag_i+c_i\qquad a_{2i}=i(c_i-c^\dag_i)\, .
\ee
We call $\Gamma$ the correlation matrix
\be
\Gamma_{i j}=\delta_{i j}-\braket{\Psi|a_i a_j|\Psi}\, .
\ee
The probability of measuring $\mu[n_j]=\frac{1+s}{2}$, with $s=\pm 1$,  is $p=\frac{1-i s\Gamma_{2j,2j-1}}{2}$. 
After the measurement  the correlation matrix transforms as follows
\be\label{eq:GammaM}
\Gamma_{\ell n}\xrightarrow{2\mu[n_j]=1+s}\Bigl[1-\frac{(\delta_{\lfloor\frac{n+1}{2}\rfloor j}-\delta_{\lfloor\frac{\ell+1}{2}\rfloor j})^2}{1- is \Gamma_{2j,2j-1}}\Bigr]\Gamma_{\ell n}+  is\frac{\delta_{2j,\ell}\delta_{2j-1,n}-\delta_{2j-1,\ell}\delta_{2j,n}+\Gamma_{n,2j-1}\Gamma_{\ell, 2j}-\Gamma_{\ell,2j-1}\Gamma_{n,2j}}{1- i s\Gamma_{2j,2j-1}}\, .
\ee

Since $[n_j,n_{j'}]=0$, all the occupation numbers in a subsystem $A$ ($j\in A$) can be measured simultaneously (this is equivalent to measuring an observable $\sum_{j\in A} \alpha_j n_j$, with incommensurate $\alpha_j$).  The resulting correlation matrix can be obtained applying several times the transformation \eqref{eq:GammaM}. The probability of the outcome is obtained analogously. 
Fig. \ref{f:mzsM} shows the time evolution of the staggered magnetization for a particular set of results of the measurements, generated randomly with the respective probabilities. The reversibility properties of the dynamics along with the regularity at which the measurements are performed and the switch is flipped can be used to reconstruct a probability distribution for the staggered magnetization. As far as expectation values are concerned, our preliminary analysis suggests that it is as if the quantum state were almost perfectly replicated before each measurement. In particular, the two phases can be distinguished by the time average of the outcomes.   

Finally, we note that a measurement of $n_{r+1} - n_{r}$ breaks gaussianity when the result is $0$, indeed the projector on the corresponding eigenspace, \emph{i.e.} $1-n_r-n_{r+1}+2n_r n_{r+1}$, is not gaussian (in terms of the fermions \eqref{eq:a}).

\section{Further Examples}\label{s:examples}
In this section we show some numerical results for the following examples:
\begin{enumerate}
\item \label{eq:OP} Switch of the boundary conditions in a fermionic XY chain in a transverse field;
\item \label{eq:NI} Switch associated with a localized magnetic field that breaks a $U(1)$ symmetry  in a nonintegrable spin-$\frac{1}{2}$ model.
\end{enumerate}
\subsection{Case \ref{eq:OP}}
\begin{figure}[!tb]
\includegraphics[width=0.45\textwidth]{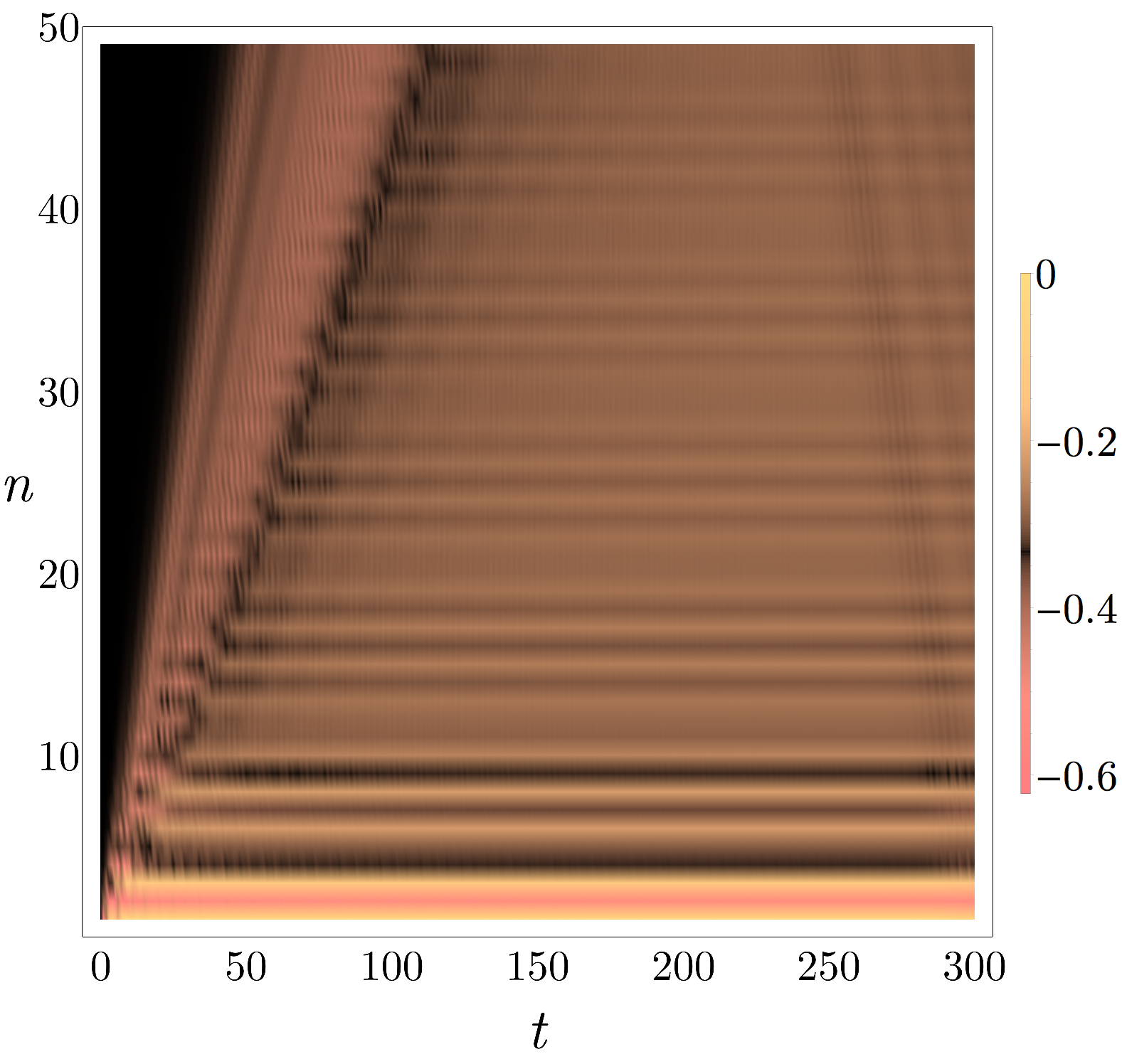}\qquad
\includegraphics[width=0.45\textwidth]{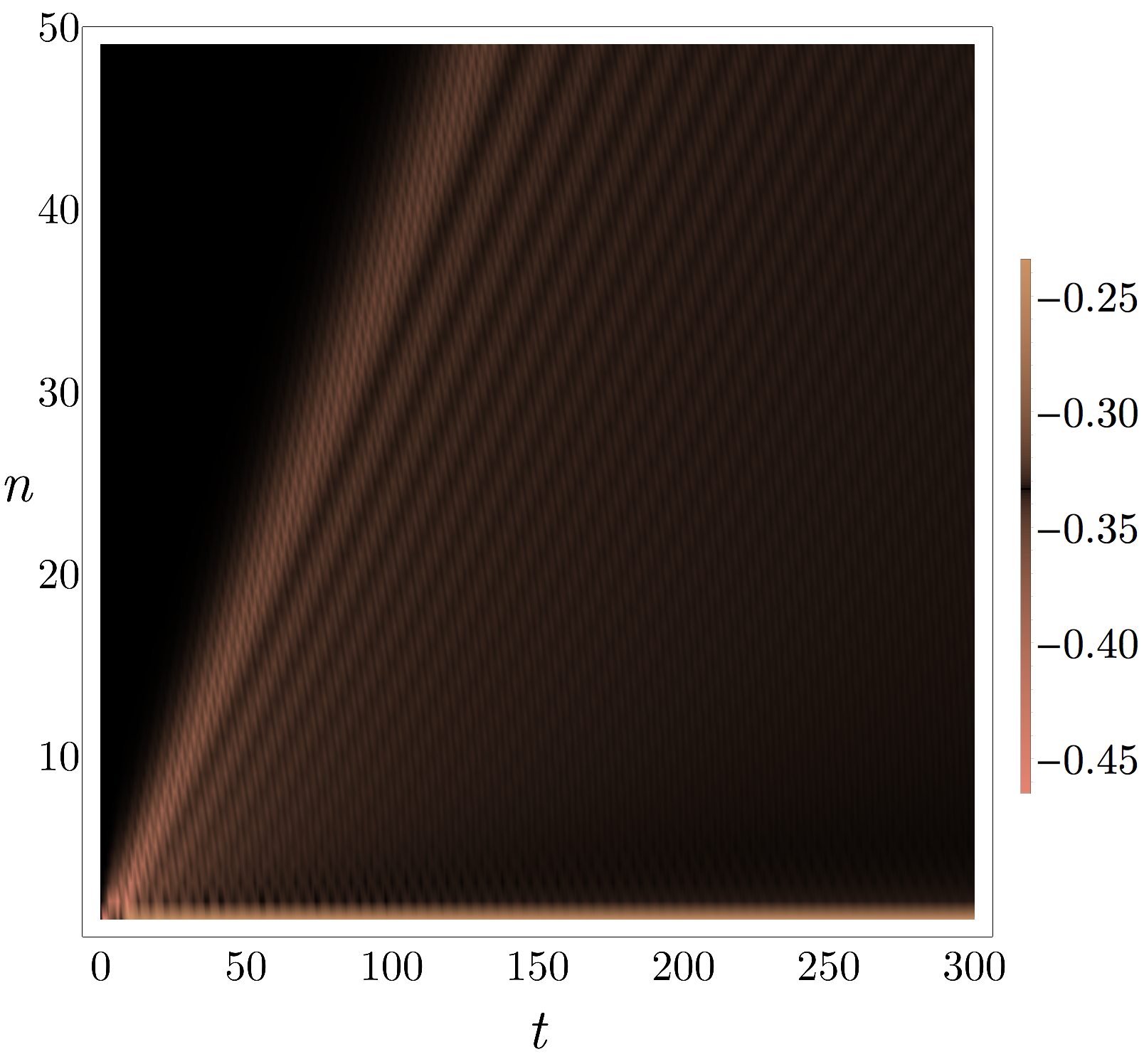}
\caption{The energy density at position $n$ and time $t$ corresponding to the time evolution of the state \eqref{eq:vacuum} under the XY Hamiltonian $H^{\rm XY}_{1}$ \eqref{eq:Hh} with $h=0.333$ and $\gamma=1.732$ (left) or $\gamma=1.1$ (right) in a chain of $300$ spins (notice that the scale of the colors is the same).
Left. There are extinct local reflection symmetric charges. In the limit $1\ll vt\ll n$ the energy density is undistinguishable from the one corresponding to time evolution with $H_0$, which is also the initial value (black); 
in the limit $1\ll n\ll vt$ a new stationary behavior emerges. 
At times $t\gtrsim 250$ finite size effects become visible.
Right. All the local reflection symmetric charges survive the boundary. The effect of the boundary fades away at large time and distance: the switch can not be used to control global properties.  
\label{f:XYOBC}
}
\end{figure}
\begin{figure}[!tb]
\includegraphics[width=0.45\textwidth]{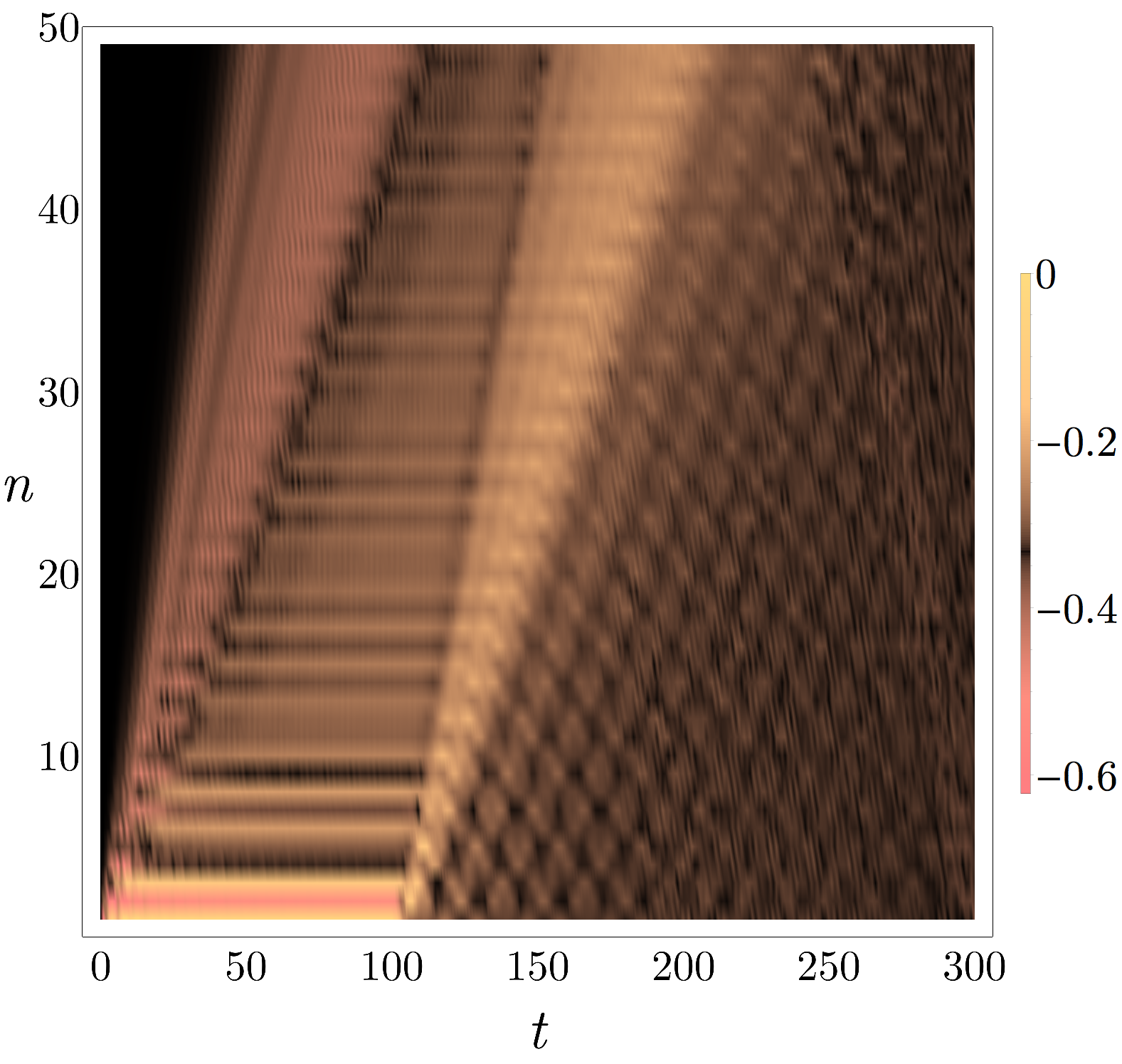}\qquad
\includegraphics[width=0.45\textwidth]{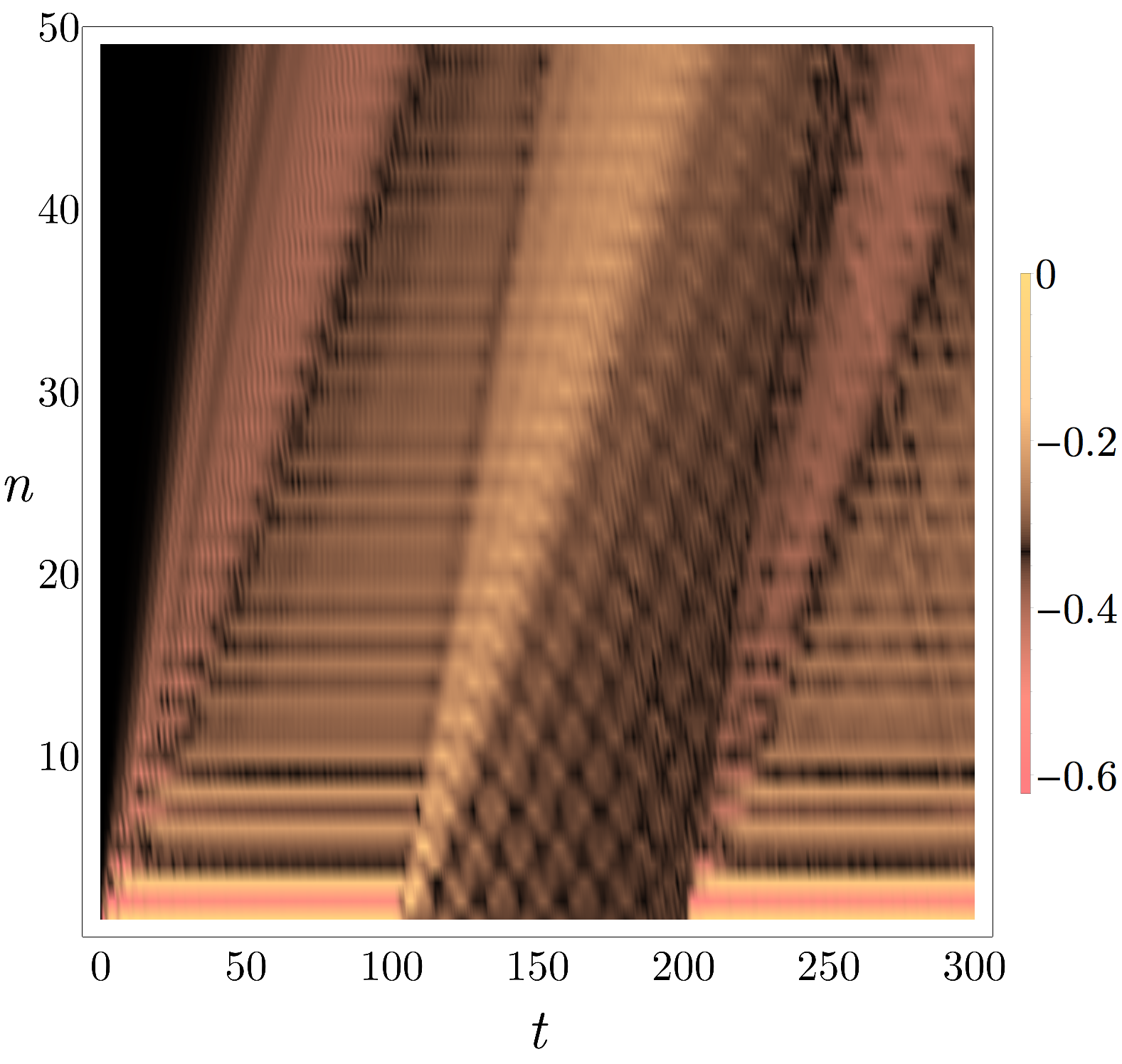}
\caption{The energy density at position $n$ and time $t$ corresponding to the time evolution of the state \eqref{eq:vacuum} under the XY Hamiltonian $H^{\rm XY}_{\mathtt s}$ \eqref{eq:Hh} with $h=0.333$ and $\gamma=1.732$ in a chain of $300$ spins. The switch is ON ($\mathtt{s}=1$) until $t=100$, then it is turned OFF ($\mathtt{s}=0$). In the right panel the switch is turned on again at $t=200$.
Left.  After flipping the switch the original value of the energy density is slowly restored. Right. After flipping the switch the second time a new light-cone emerges with the same morphology of the original one.  
\label{f:XYOBCswitch}
}
\end{figure}

Let us include a magnetic field $h$ to the Hamiltonian \eqref{eq:H}
\be\label{eq:Hh}
H^{\rm XY}_{\mathtt s}=\sum_{\ell=1}^L\Bigl[\frac{1}{2}(c_\ell c^\dag_{\ell+1}-c^\dag_\ell c_{\ell+1})+\frac{\gamma}{2}(c_\ell c_{\ell+1}-c^\dag_\ell c^\dag_{\ell+1})+\frac{h}{2} (2c_\ell^\dag c_\ell-1)\Bigr]-\mathtt{s}\Bigl[\frac{1}{2}(c_L c^\dag_{1}-c^\dag_L c_{1})+\frac{\gamma}{2}(c_L c_{1}-c^\dag_L c^\dag_{1})\Bigr]\, ,
\ee
where we imposed periodic boundary conditions $c_{L+1}=c_1$.
The field destroys all the extra (non-abelian) charges that characterize the XY model in zero field when $L$ is even. The switch changes the boundary conditions from periodic ($\mathtt s=0$) to open ($\mathtt s=1$).

As initial state we take the fermion vacuum
\be\label{eq:vacuum}
c_\ell\ket{\Psi_0}=0\, .
\ee
We set $h=0.333$ and consider two values for the anisotropy: $\gamma=1.732$ and $\gamma=1.1$. 

Generally, in integrable models, changing the boundary conditions from periodic to open destroys the charges that are odd under chain inversion~\cite{GM:open}. 
However, the initial state is reflection symmetric, so the extinction of the odd charges should not be relevant to time evolution. 
This example is still significant because Ref.~\cite{F:XYopen} showed that the set of (quasi)local conservation laws for $\mathtt s=1$ depends on the ratio $q=|h/(\gamma^2-1)|$; specifically, for $q>1$ all the reflection symmetric charges of the periodic chain survive the boundary, for $q<1$ instead there are less conservation laws. 
The two values of $\gamma$ that we consider are representative of the two situations. Consequently, for $\gamma=1.732$ the switch is expected to control global properties, while for $\gamma=1.1$ it is not.

Fig.~\ref{f:XYOBC} shows the main difference between the case $q<1$ and $q>1$. In the former, time evolution under $H_0$ results in a stationary state different from the one with periodic boundary conditions ($\mathtt s=0$).  On the other hand, for $q>1$ the limit of large time seems to commute with the limit of large distance. This is perfectly consistent with the theoretical picture presented in Sec.~\ref{s:gen}. 

In Fig.~\ref{f:XYOBCswitch} we check reversibility by turning off the switch and flipping it again after a sufficiently large time. The novel light-cone has the same morphology of the original one.

We would like to point out that the particular profile of the energy density is probably model dependent and, for example, the rather complicated behavior in the limit of infinite time at fixed ratio $\frac{n}{t}$ could be a consequence of integrability or of the absence of interactions.  
Nevertheless, as shown in Sec.~\ref{s:gen}, the noncommutation of the large time limit with the large distance turns out to be a general property of any model with a defect that destroys a relevant set of conservation laws. 

\subsection{Case \ref{eq:NI}}
\begin{figure}[!tb]
\includegraphics[width=0.75\textwidth]{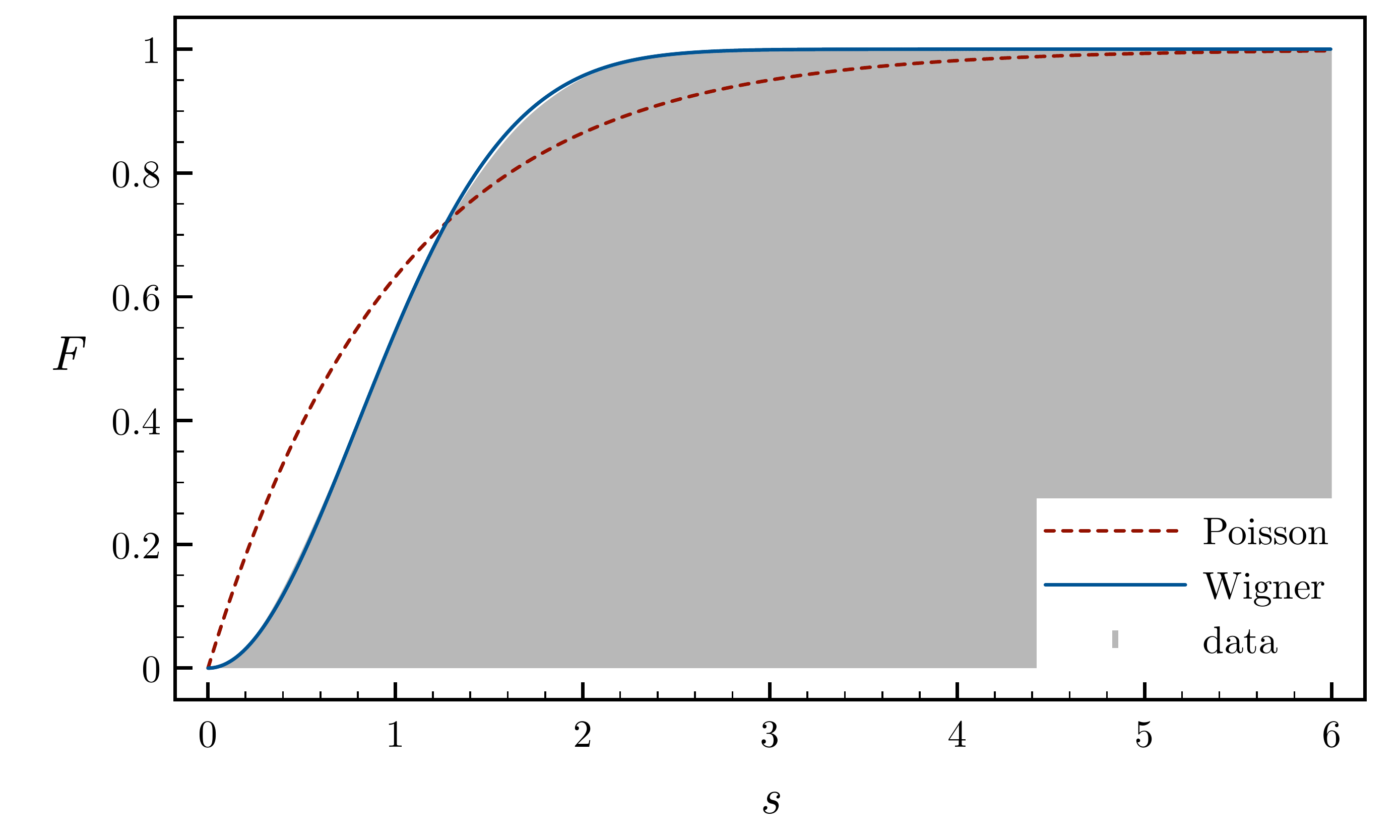}
\caption{The cumulative distribution function $F(s)$ of the nearest neighbor spacing distribution (\emph{i.e.} the probability density of the difference between two consecutive levels of the unfolded spectrum of the energy) for the Hamiltonian $H_0$ \eqref{eq:HNI} with  $\Delta=0.5$, $g=-0.5$, $U=0.25$ in the reflection symmetric sector with zero momentum, zero magnetization and spin flip $\prod_\ell\sigma_\ell^z\rightarrow 1$ for $L=20$. The data are very well described by a Wigner distribution ($F(s)=1 - e^{-\frac{\pi s^2}{4}}$), which is typical of nonintegrable models (in integrable models the typical distribution is instead Poisson $F(s)=1-e^{-s}$).  
\label{f:Wigner}
}
\end{figure}

\begin{figure}[!tb]
\includegraphics[width=0.75\textwidth]{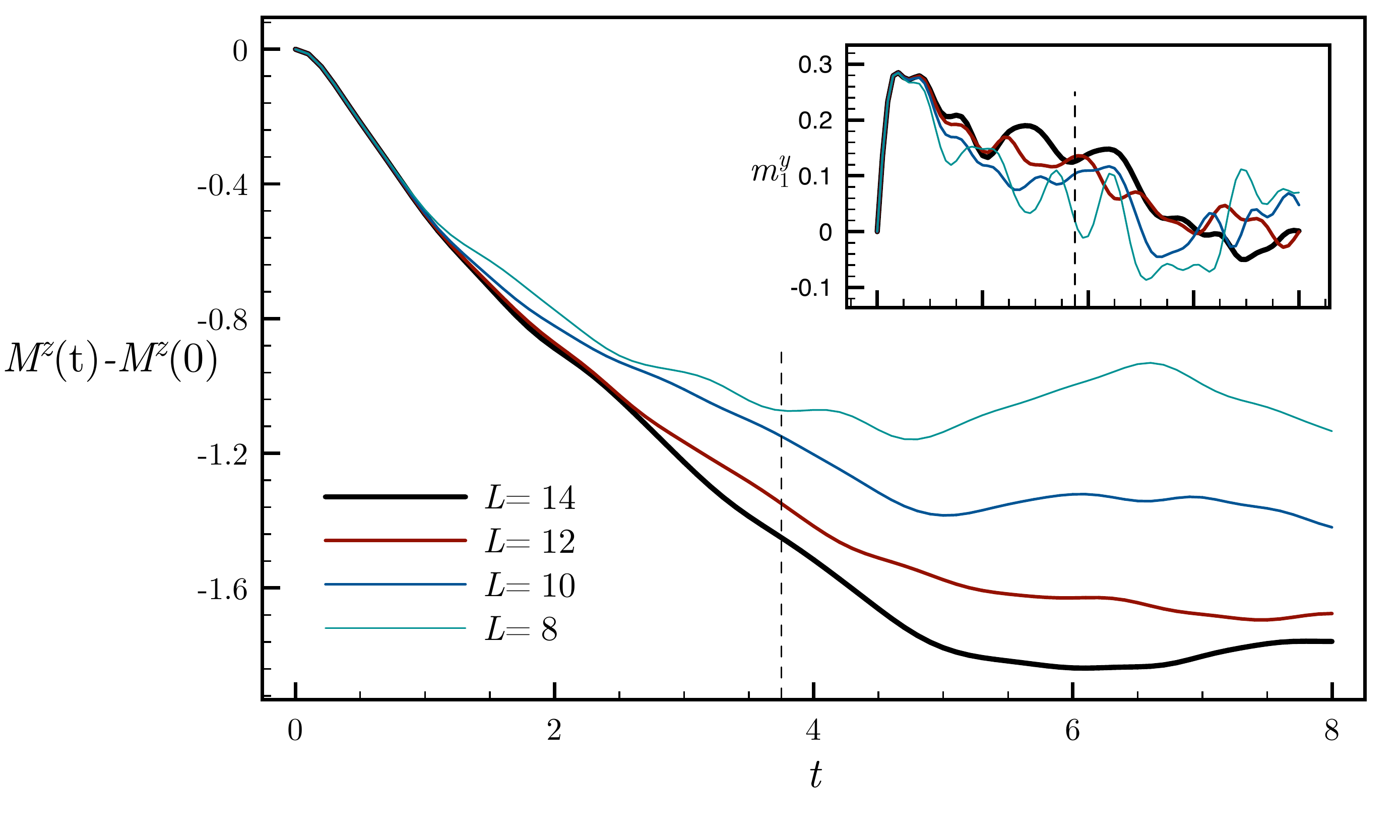}
\caption{The variation in the total magnetization $M^z=\frac{1}{2}\sum_\ell\braket{\sigma_\ell^z}$ after a quench from the state~\eqref{eq:Psi0} under Hamiltonian~\eqref{eq:HNI} with $\Delta=0.5$, $g=-0.5$, $U=0.25$ and $\mathtt s=1$, for several chain sizes. 
The vertical dashed line is an estimate of the time at which finite size effects become significant even for $L=14$.
Data are consistent with a linear decrease of $M^z$. In the inset, the local magnetization $m_1^y=\frac{1}{2}\braket{\sigma_1^y}$, which is proportional to the time derivative of $M^z$.
\label{f:nonint}
}
\end{figure}
\begin{figure}[!tb]
\includegraphics[width=0.95\textwidth]{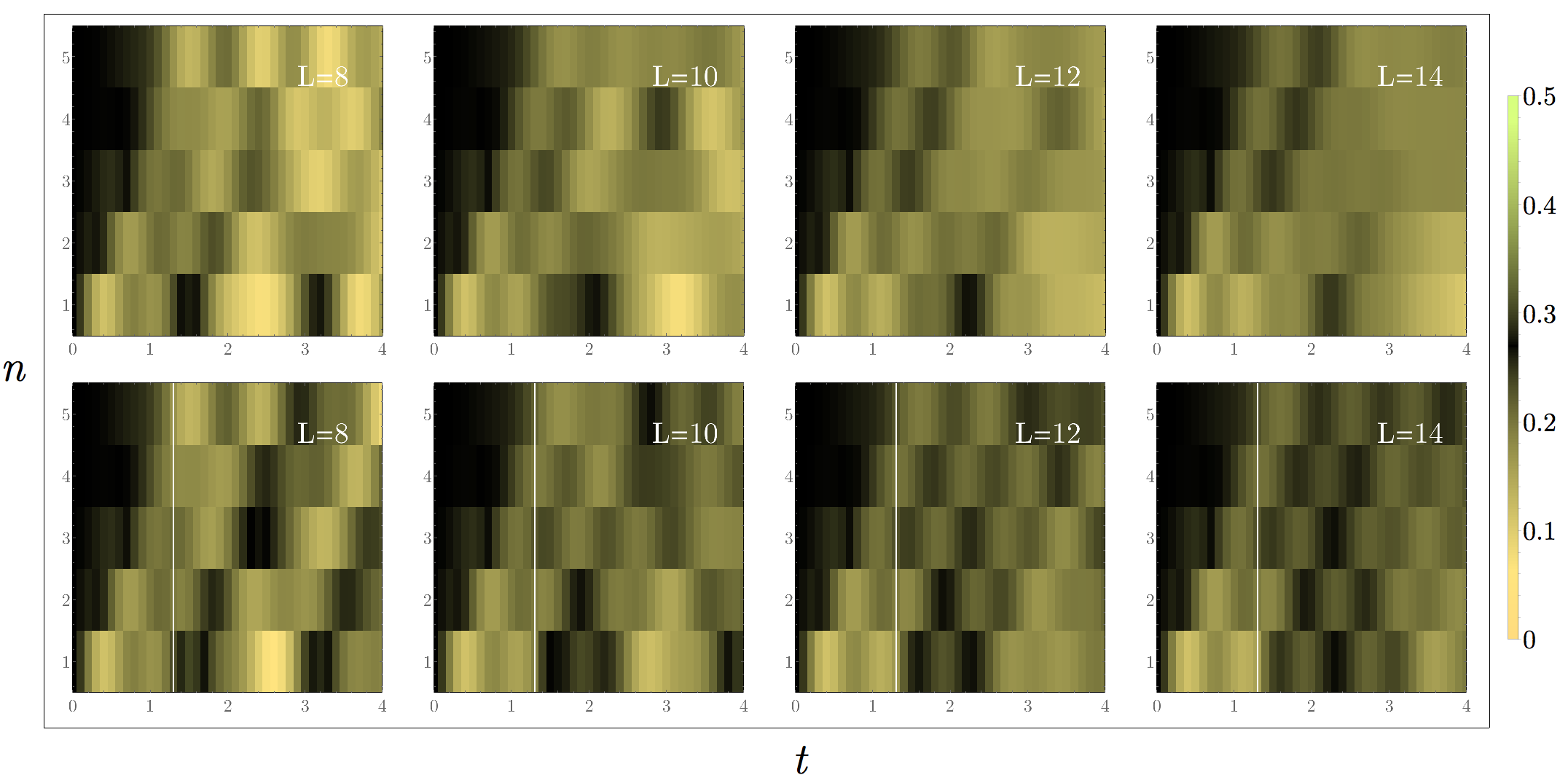}
\caption{The profile of the local magnetization $m^z_n=\frac{1}{2}\braket{\sigma_n^z}$ after a quench from the state~\eqref{eq:Psi0} under Hamiltonian~\eqref{eq:HNI} with $\Delta=0.5$, $g=-0.5$, $U=0.25$ for the chain sizes $L=8,10,12,14$. In the upper panels the switch is always ON ($\mathtt{s}=1$); in the lower panels the switch is turned off at $t=1.3$. 
\label{f:nonintpan}
}
\end{figure}

\begin{figure}[!tb]
\includegraphics[width=0.95\textwidth]{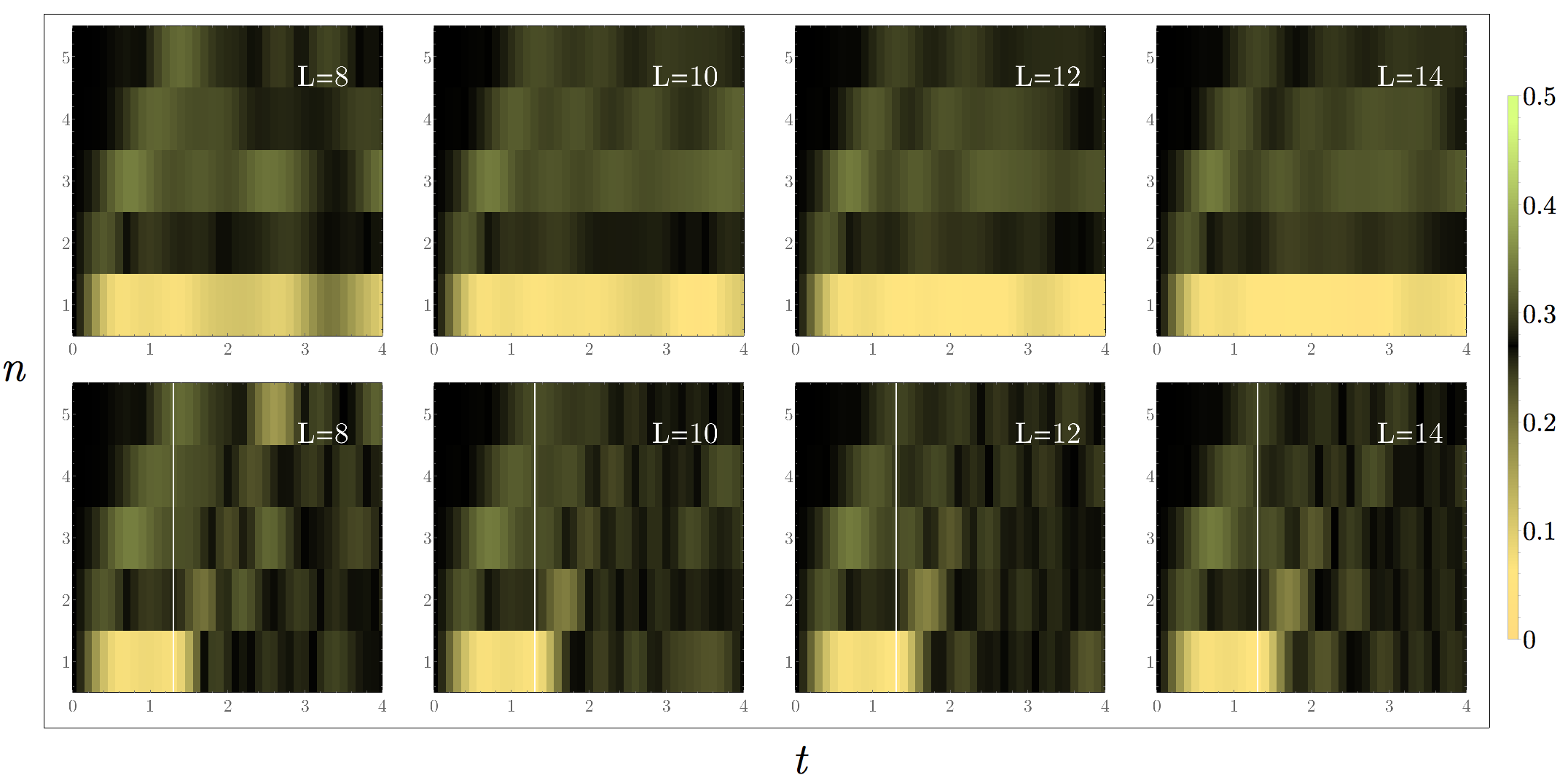}
\caption{The profile of the local magnetization $m^z_n=\frac{1}{2}\braket{\sigma_n^z}$ after a quench from the state~\eqref{eq:Psi0} under Hamiltonian~\eqref{eq:H2} with $\Delta=0.5$, $g=-0.5$, $U=0.25$ for the chain sizes $L=8,10,12,14$. In the upper panels the switch is always ON ($\mathtt{s}=1$); in the lower panels the switch is turned off at $t=1.3$. 
\label{f:nonintpanz}
}
\end{figure}
As a final example, we consider a nonintegrable model with Hamiltonian
\be\label{eq:HNI}
H_{\mathtt{s}}=\sum_{\ell=1}^L \Bigr[\sigma_\ell^x\sigma_{\ell+1}^x+\sigma_\ell^y\sigma_{\ell+1}^y+\Delta \sigma_\ell^z\sigma_{\ell+1}^z+g\sigma_\ell^z\sigma_{\ell+2}^z+U(\sigma_\ell^x\sigma_{\ell+1}^z\sigma_{\ell+2}^x+\sigma_\ell^y\sigma_{\ell+1}^z\sigma_{\ell+2}^y)\Bigl]-\mathtt{s}\sigma_1^x\, ,
\ee
where periodic boundary conditions $\sigma_{L+1}^\alpha=\sigma_1^\alpha$ are imposed and $\mathtt{s}$ takes the value $1$ or $0$ depending on whether the switch is ON or OFF.
We set the Hamiltonian parameters as follows: $\Delta=0.5$, $g=-0.5$ and $U=0.25$.
Fig.~\ref{f:Wigner} shows that this choice makes the model generic. 
For $\mathtt s=0$ the Hamiltonian is invariant under rotations about $z$, indeed it commutes with $S^z=\frac{1}{2}\sum_\ell\sigma_\ell^z$. 

In the following we provide numerical evidence that the local magnetic field in the $x$ direction, which breaks rotational symmetry around $z$, is responsible for the emergence of a front that separates two regions with different magnetization. 

As initial state we take the product state
\be\label{eq:Psi0}
\ket{\Psi_0}=e^{\frac{i}{2}\sum_{\ell}\sigma_\ell^y}\ket{\uparrow\cdots\uparrow}
\ee
We study time evolution by exact diagonalization techniques, which allow us to access only small systems (we investigated chains up to $L=14$). 

The total magnetization $S^z$ does not commute with $H_1$.
A generic defect that does not commute with $S^z$ is expected to destroy the conservation law $S^z$. 
As discussed in Sec.~\ref{s:gen}, the standard signature would be the presence of a nonzero slope in the time evolution of the total magnetization in the $z$ direction.
Although the finite size effects are significant, Fig.~\ref{f:nonint} is compatible with a linear decrease in time (which should be regarded as  the standard situation) and hence we deduce that there is \emph{no} term (quasi)localized around the defect that makes $S^z$ commuting with $H_1$. 

The upper panels of Fig.~\ref{f:nonintpan} show the profile of the local magnetization $m_\ell^z=\frac{1}{2}\braket{\sigma_\ell^z}$ as a function of the time for various system sizes (to give some idea of the finite size effects). One can fairly notice the emergence of a macroscopic region with a different local magnetization. The lower panels correspond to turning off the switch at the time $t=1.3$.  One can perceive the propagation of a new light-cone that is going to restore the original magnetization.
The dimensions of the systems investigated are clearly too small to deduce any behavior purely from numerics, however the data are consistent with the development of regions with globally different properties, as explained in Sec.~\ref{s:gen}.

Fig.~\ref{f:nonintpanz} shows the results of the same analysis for a different switch that turns on a local magnetic field in the $z$ direction
\be\label{eq:H2}
H_2=H_0-\sigma_1^z\, .
\ee
This time the Hamiltonian remains invariant under rotations about $z$ and we do not expect macroscopic effects. Fig.~\ref{f:nonintpanz} is compatible with our belief, in the sense that the behavior inside the light-cone appears closer to the one for $\mathtt s=0$ (the initial value). 

\end{document}